# Study of relativistic charged particles production in $^{84}Kr_{36}$+Em interaction at around 1 A GeV with wounded nucleon model

N Marimuthu[1, 2], V Singh[1]*, S S R Inbanathan[2]*
[1]*Department of Physics, Institute of Science, Banaras Hindu University, Varanasi - 221005, INDIA*
[2]*Post Graduate and Research Department of Physics, The American College, Madurai-625002, TN, INDIA*
*E-mail: *venkaz@yahoo.com, stepheninbanathan@gmail.com*

**Abstract:** This article is focused on the multiplicity and probability distribution of the emitted charged pions ($N_\pi$) for interaction of $^{84}Kr_{36}$ projectile with nuclear emulsion targets at kinetic energy 1 GeV per nucleon. In the wounded nucleon model frame work, we have calculated the total number of wounded nucleons (W) and total number of interactions (ν). The obtained results revealed that the average multiplicity of the charged pions $\langle N_\pi \rangle$ is dependent on the projectile and target mass. The calculated values of the wounded nucleons (W) and total number of interactions (ν) shows strong dependence on the mass of the colliding nuclei. The emission rate of mean multiplicity of the $\langle N_s \rangle$ and $\langle N_\pi \rangle$ increases with increasing the total number of wounded nucleons (W) and total number of interactions (ν). The mean multiplicity value of $\langle N_s \rangle$ and $\langle N_\pi \rangle$ linearly increases with increasing the projectile mass number ($A_P$) and independent of the incident energy.

**Keywords:** Nuclear Emulsion detector; Multiplicity distribution; wounded nucleon model.

**PACS No.:** 25.75.-q, 29.40.Rg, 25.70.Mn.

## 1. Introduction

The study of relativistic nucleus – nucleus (A-A) and hadrons – nucleus (h-A) collisions have progressed in the recent years and such collisions may reveal the long awaited new phase transition called Quark-Gluon Plasma (QGP) [1-4]. The heavy-ion collisions are providing information on the deep properties of nuclear matter, which is very useful in the investigation of cardinal equation in the nuclear and sub-nuclear physics. The various concepts of nuclear physics are extrapolated with heavy-ion collisions and it is essential to check their application in wide range of incident energies i.e. 1 to 200 GeV/n [5-6]. According to the participant spectator model (PS Model) [7-9], the interacting systems are divided into three different regions: projectile spectator (PS), target spectator (TS) and participant region i.e. fire-ball. The overlapping region of the nuclear volumes are called the participant region, where the nuclear matter violently interacts with each other and create a large number of new particles depending on various factors. The newly produced particles are mostly mesons (π-mesons), photons, lepton pairs etc. The remaining parts of nuclei, which do not participate directly in the collision, creates various charged and neutral fragments through the disintegration process, are called spectators. On the other hand, Nuclear Emulsion Detector (NED) has played prominent role in the studies of high energy interactions during the twentieth century. The beauty of NED is to provide excellent particle detection with unprecedented position resolution i.e. better than 1μm, over the full solid angle i.e., 4π geometry coverage [10]. Pions production is one of the most dominating processes in the high energy heavy-ion collisions. At the projectile energy in between 1 to 2 GeV per nucleon range, the pions to participant ratio is about 10-20% [11]. If the projectile energy is in the range of 5-10 GeV per nucleon, the pions to participant ratio reaches up 70-100%. In the ultra relativistic high energy regions (i.e. >100 GeV per nucleon) pions are the most abundant particles [11]. From the long period of time, the study of produced pions are intensely investigated in the wide range of incident energies because of high energy pions are considered to be precise tool to understand the properties of the hot and dense nuclear matter produced in the participant region during the collisions [12]. Moreover, in the heavy-ion collisions most of the energy is spent in the creation of particles, particularly the pions. To shed light on the collision dynamics and properties, and basic reaction mechanism of such collision, it is necessary to understand the pion production mechanism [13]. In this work, the number of produced charged pions ($N_\pi$) is obtained from the produced shower particles ($N_s$) and projectile spectator proton ($N_p$) i.e. $N_\pi = N_s - N_p$, during the interactions. The presented results are obtained from an experiment, where nuclear emulsion plates are exposed with $^{84}Kr_{36}$ beam having Kinetic Energy (KE) 1 GeV per nucleon. We have investigated the multiplicity and probability distribution of produced charged pion i.e. $\langle N_\pi \rangle$ and obtained results are compared with the others results. We also investigated the dependence of average number of produced relativistic charged particles on the total number of the interactions and the wounded nucleons calculated using the wounded nucleon model. Such study will shed light on understanding of the particle productions and nucleus – nucleus (A-A) interactions.

## 2. Experimental Details

In this article, analyzed data were collected from the nuclear emulsion detector experiment in which employed a stack of highly sensitive NIKFI-BR2 emulsion plates with 600 μm unprocessed thickness. These emulsion plates irradiated horizontally at in Gesellschaft fur Schwerionenforschung (GSI), Darmstadt (Germany) by the $^{84}Kr_{36}$ beam with Kinetic Energy (KE) around 1 GeV per nucleon [9]. In order to obtain the primary interactions, along the primary track scanning technique has been adopted using Olympus BR-2 Binocular microscope. Generally two types of scanning methods have been used to study the emulsion plates, first one is line scanning and second one is volume scanning. Here interaction is found through the line scanning method with oil immersion objective lens of 100× along with the 15× eyepieces. In case of the line scanning method, beam tracks are picked up at 5mm distance from the edge of the emulsion plate and followed very carefully, till they are interacted with the nuclear emulsion nuclei or stopped in the plate or escape from any surface of the plate. To ensure that we are following the primary track, first we followed the primary track in the backward direction until the edge of emulsion plates [10]. In case of the volume scanning method, emulsion plate's information is collected through the strip-by-strip scanning. The emitted charged particles from the primary interaction are called secondary charged particles and are divided into the following categories,

**Black particles** ($N_b$) - Black particles are slow evaporated target fragments having kinetic energy and range less than 20 MeV and 3 mm, respectively. These particles are mostly coming from the excited target nuclei or fragments i.e. target spectator of the interaction.

**Grey particles** ($N_g$) - Grey particles are fast target fragments or nucleons, mainly the recoil protons, having kinetic energy and relativistic velocity ($\beta$) in between 30 to 400 MeV and 0.3c to 0.7c, respectively.

Here black and grey particles combined together are called the heavily-ionizing charged particles and it is denoted by $N_h = (N_b + N_g)$. It means $N_h$ is almost representing the number of target fragments.

**Shower particles (**$N_s$**)** - Shower particles are singly charged relativistic particles having velocity greater than 0.7c. These particles are mostly produced in the fireball or participant region during the interactions. The number of such particle represents the degree of destruction, and depends on collision geometry and impact parameter (b) of the event.

**Projectile fragments** – Projectile fragments are categorized according to their charges that is singly ($N_p$), doubly ($N_\alpha$) and multiply ($N_f$) charged projectile fragments. Projectile fragments mostly belong to the projectile spectator region.

### 3. Wounded nucleon model

Wounded nucleon model has been proposed mainly to understand the multiparticles production in nucleus – nucleus (A-A) collision at high energies. The wounded nucleon i.e. participant nucleon is one of the primary tools for explaining and giving the interpretation of phenomena occurred in high energy heavy-ion collisions. According to the predictions of the "wounded" nucleon model, the number of produced relativistic charged particle multiplicities ($n_{AA}$) in the nucleus – nucleus (A-A) collision show the scaling property of the average number of participant or wounded nucleons (W) [11, 14]

$$n_{AA}(E) = \tfrac{1}{2} W n_{PP}(E). \qquad (1)$$

Here, $n_{PP}(E)$ represents the proton - proton particle multiplicity at an equivalent energy. Another convenient and useful parameter to calculate the multiplicity per participating nucleon is given as M = n / W. Here, the wounded nucleon W contains the all quantities of geometrical effects such as nuclear radius, density and impact parameter. It is believed that M depends only on the interaction dynamics of the collision and not on the impact parameter (b) [15-16]. From the purely geometrical aspects, the number of wounded nucleons or participating nucleons from the projectile and target nuclei, and the nuclear interactions are given in terms of the interaction cross-section as

$$W = W_P + W_T = \frac{A_P \sigma_{PA_T}}{\sigma_{A_P A_T}} + \frac{A_T \sigma_{PA_P}}{\sigma_{A_P A_T}}. \qquad (2)$$

And the total number of interactions is given by

$$\nu = W_P \nu_T = W_T \nu_P. \qquad (3)$$

Where

$$\nu_T = \frac{A_T \sigma_{NN}}{\sigma_{PA_T}}. \qquad (4)$$

Here, $W_P$ and $W_T$ are the number of wounded projectile and target nucleons, and $\nu_P$ and $\nu_T$ are the average number of projectile and target collisions. In equation (4), nucleus - nucleus ($\sigma_{A_P A_T}$), nucleon - nucleus ($\sigma_{PA}$) and nucleon - nucleon ($\sigma_{NN}$) cross-sections are calculated from equations described in Refs. [17-18].

**4. Results and Discussion**

For this analysis, 615 primary inelastic events were collected from the $^{84}Kr_{36}$ beam interacted with emulsion detector's nuclei with KE around 1 GeV per nucleon. The nuclear emulsion detector mostly contains hydrogen (H), carbon (C), nitrogen (N), oxygen (O), sulfur (S), bromine (Br), silver (Ag) and iodine (I). The exact target identification on the event-by-event basis is not possible in the NED experiments. Although, various methods have been proposed and used to identify the target involved in an interaction of both nucleus – nucleus (A-A) and hadrons – nucleus (h-A) collisions. However statistically, we have identified the emulsion target nuclei involved in the interactions on the basis of heavily-ionizing charged particles i.e. $N_h$ [10]. As we know that $N_h$ basically belongs to the target fragments and that implies the size of the target nucleus. Therefore, the events with $N_h$ equal to 0 and 1 are considered as events belong to the Hydrogen (H) target. An interaction having $N_h$ more than 7 is definitely belongs to the Ag or Br target interaction. While events with $N_h$ value in between 2 to 7 are due to the C, or N, or O target interactions. It has been estimated through observation that the numbers of corresponding events due to the H, CNO and Ag / Br target groups were 10.5%, 23.9% and 65.5%, respectively. The Multiplicity distribution of the produced pions ($N_\pi$) in $^{84}Kr_{36}$ - Emulsion interactions at around 1 A GeV is shown in the figure 1. It is evident from figure 1 that the distribution shapes of the emitted pions are varying with the target groups of the emulsion nuclei and at the same time the distributions are getting wider with increasing of target size. It is also interesting to note that the number of the pions ($N_\pi$) emitted from heavier target nuclei i.e. Ag / Br are comparatively higher than the lower mass target such as H, CNO nuclei in the emulsion. It implies that pion production is more favorable in case of symmetric collisions where projectile and target have similar masses. The maximum number of emitted pions $(N_\pi)_{max}$ is 19, 20, and 30 in case of H, CNO and Ag / Br target groups, respectively.

Table 1 illustrates the mean multiplicity of produced pion $\langle N_\pi \rangle$ of $^{84}Kr_{36}$ interactions with H, CNO and Ag / Br target groups at 1 GeV/n along with the values of other systems at different energies. From the table one can easily observe that the mean multiplicity of the emitted pions $\langle N_\pi \rangle$ for different target groups of emulsion nuclei increases with increasing target masses, however, at the same multiplicity values is almost constant within error with respect to the total kinetic energy of the projectile except hydrogen projectile.

The probability distribution of the emitted pions $<N_\pi>$ is plotted in the figure 2 and the distribution is fitted with the Gaussian function. The best fitting parameters of the Gaussian distribution function is -5.14 and 26.88, for the peak and width, respectively. It is important to mention here that the various target regions are marked in figure 2. The maximum value of the probability distribution for pions production is found to be 15.13±1.36.

In table 2, we have compared the average multiplicities of the produced charged pions ($<N_\pi>$) emitted in the present experiment and compared with different projectiles having different total kinetic energy but almost similar kinetic energy per nucleon. From table 2, one can see that the average multiplicities of emitted pions $<N_\pi>$ increases with increasing projectile mass number i.e. with increasing total kinetic energy of the projectile.

We also investigated the dependence of $<N_\pi>$ on the average target mass number ($A_T$) in $^{84}$Kr - Emulsion interactions at around 1 GeV per nucleon and compared with the different projectiles as shown in figure 3. It may be seen from figure 3 that the average $<N_\pi>$ values increases with increasing projectile and target mass number. The observations are fitted with Eq. (5),

$$<N_\pi> = a\,(A_T)^b. \qquad (5)$$

Where, $A_T$ represents the average target mass number of different target groups based on the number of heavily ionizing charge particle $N_h$ in $^{84}$Kr - Emulsion interactions. The best fitting parameters are a = 0.20±0.06 and b = 0.37±0.04 for $^{84}$Kr - Emulsion interactions. Comparing the corresponding fitting parameters for the $^{24}$Mg (a = -0.06±0.03, b = 0.49±0.02) and $^{28}$Si (a = - 0.26±0.05, b = 0.63±0.04) [11, 17].

We have used the wounded nucleon model discussed in the section 3 for calculation of total number of the wounded nucleons (W) or participating nucleons, and the total number of interactions (ν) for different projectiles such as $^{84}$Kr-Em**,** $^{132}$Xe-Em**,** and $^{197}$Au-Em at incident kinetic energy around 1 GeV per nucleon. In addition, we have also calculated the same for projectile $^{56}$Fe-Em at incident energy 1.8 A GeV. Table 3 shows that the calculated values of W and ν as well as observed values of the their average number of shower $<N_s>$ particles for inelastic collisions of different projectiles at different kinetic energies, whereas calculated values are compared with other projectiles at different energies. From the table 3, one can observe that the total number of wounded nucleons (W) and total number of interactions (ν) are both substantially increase with increasing the different target nuclei. In addition, one can observe that, the total number of wounded nucleons and interactions are also increases with increasing colliding nuclei. It is clear that the production of both W and ν are not only depending on the target mass number but also strongly depending on the mass number of the projectile nuclei.

Usually studies are based on the projectile energy per nucleon however, total energy of the projectile is reveling more information. Therefore, a distribution between mean numbers of produced pions is plotted as a function of total energy for different projectiles in figure 4. From figure 4, it is clear that the production mechanism is strongly dependent on the total energy of the projectile.

Figure 5 depicts a linear relationship between the averages number of shower particles $<N_s>$ and wounded nucleons (W) for interaction of the various projectiles at different incident energies ranging between 1 to 1.8 and 2.2 to 3.3 A GeV. It is important to note that in both energy regions, the production rate of average shower particles, which is a mixture of pions, kaons and other singly charge particles, are significantly different and it increases with increasing incident kinetic energy.

Similarly, we have also plotted for the emitted pions $<N_\pi>$ as a function of the calculated number of wounded nucleons (W) in figure 6. In light of figure 5, one can see from figure 6 that at similar energy range the average production rate of shower particle ($<N_s>$) is 0.29±0.02, which is greater than the average production rate 0.19±0.01 of pions ($<N_\pi>$). The number of showers produced in an event is more than the number of pions estimated, which is obvious and it is evident. It is also evident from both figures 5&6 that the shower and pion production rate with respect to wounded nucleons (W) seems to be strongly dependent on the incident projectile energy. The present experimental data points are fitted with straight line function.

$$<N_s> = (0.29\pm0.02)W + (2.25\pm0.06) \quad (6)$$

$$<N_\pi> = (0.19\pm0.01)W - (1.02\pm0.30) \quad (7)$$

Figure 7 and 8 represent the dependence of mean multiplicities of shower $<N_s>$ particles and mean pion multiplicities $<N_\pi>$ on the total number of estimated interactions or collisions (ν) between nucleons for different projectiles with emulsion nuclei in different energy regions. From the figures 7 & 8, one may see that the mean multiplicities of the charged particles i.e. $<N_s>$, and $<N_\pi>$ increases with increasing the total number of estimated interactions.

The rate of change of mean multiplicities of singly charged particles with respect to the number of estimated interactions is higher (0.35±0.01) in case of 2.2 to 3.3 A GeV energy regions than the 1.0 to 1.8 A GeV energy regions. It can be understood from the above mentioned plots that the geometry of nucleus-nucleus interactions and participating nucleons is still playing dominant role in the high energy collisions. The experimental data points fitted with straight line functions are as given below,

$$\langle N_s \rangle = (0.19 \pm 0.02)\,\nu + (3.24 \pm 0.02) \quad (8)$$

$$\langle N_\pi \rangle = (0.18 \pm 0.02)\,\nu - (0.98 \pm 1.07). \quad (9)$$

It is interesting to study the dependence of mean multiplicities of singly charged particles specially $\langle N_s \rangle$ and $\langle N_\pi \rangle$ as shown in the figure 9 and 10, at the mass number of the incident projectiles ($A_P$) in nuclear emulsion interactions. From figures 9 & 10, one can observed that the mean multiplicity of produced singly charged particles especially $\langle N_s \rangle$ and $\langle N_\pi \rangle$, have linear dependence on the mass number of the incident projectiles.

It may also be observed that the rate of mean multiplicity of singly charged produced particles with respect to the atomic mass number of projectile is independent of the incident projectile energy. The solid lines are showing the linear fit of the experimental data points as in equation:

$$\langle N_s \rangle = (7.01 \pm 0.62)\,A_P + (0.35 \pm 0.85). \quad (10)$$

$$\langle N_\pi \rangle = (4.80 \pm 0.50)\,A_P - (0.27 \pm 0.62). \quad (11)$$

## 5. Conclusions

In the present article, we have remarkably investigated the multiplicity distribution and probability distribution of the emitted charged pions ($\langle N_\pi \rangle$) in $^{84}Kr_{36}$ - Emulsion interaction at around 1 GeV per nucleon. From this study, we may conclude that the produced mean multiplicity of the charged pions particles $\langle N_\pi \rangle$ is strongly depending on the target mass number of the emulsion nuclei. The average multiplicities of the emitted $\langle N_\pi \rangle$ values are completely dependent on the mass number of the projectile with incident kinetic energy. We observed that the total energy of the colliding projectile has strong dependence on the production of pions. We have also calculated the total number of wounded nucleons (W) and total number of interactions ($\nu$) in the framework of wounded nucleon model. The calculated values of the wounded nucleon (W) and the total number of interactions ($\nu$) are not only depending on the target mass of emulsion nuclei but also strongly dependent on the mass of the colliding nuclei. The dependence of the average multiplicity of the charged particles i.e. $<N_s>$ and $<N_\pi>$ on the total number of wounded nucleons and total number of interactions show that the emitted relativistic charged particles $<N_s>$ and $<N_\pi>$ values are strongly depend on the total number of participating nucleons (W) and also total number of interactions ($\nu$). The correlation relation between $<N_s>$, $<N_\pi>$ and projectile mass ($A_P$) shows that, the $<N_s>$ and $<N_\pi>$ values are linearly increases with increasing the projectile mass ($A_P$) and independent of energy.

**Acknowledgment**

Authors are thankful to all the technical staff of GSI, Germany for exposing nuclear emulsion detector with $^{84}Kr_{36}$ beam and Department of Science and Technology (DST), New Delhi, for their financial support.

**Table 1:** Mean multiplicity of the charged pions <$N_\pi$> for different target groups and projectiles with different incident kinetic energy.

| Projectile | Energy (A GeV / GeV) | <$N_\pi$> | | | |
|---|---|---|---|---|---|
| | | H | CNO | Ag/Br | Reference |
| P | 3.7 / 3.7 | - | 0.68±0.31 | 0.55±0.04 | [19] |
| $^{24}$Mg | 3.7 / 88.8 | 0.92±0.11 | 2.95±0.17 | 9.08±0.31 | [11] |
| $^{28}$Si | 3.3 / 92.4 | 0.58±0.13 | 2.42±0.21 | 10.38±0.34 | [17] |
| $^{84}$Kr | 1.0 / 84.0 | 1.7±0.3 | 3.55±0.32 | 9.17±0.15 | **Present work** |

**Table 2:** The average multiplicities of produced charged pion <$N_\pi$> at various energies with different projectiles - emulsion inelastic interactions are listed.

| Collision Type | Energy (A GeV / GeV) | <$N_\pi$> | Reference |
|---|---|---|---|
| P-Em | 3.7 / 3.7 | 0.63±0.02 | [19] |
| $^{2}$H-Em | 3.7 / 7.4 | 1.5±0.1 | [20] |
| $^{3}$He-Em | 3.7 / 11.1 | 2.27±0.11 | [20, 21] |
| $^{4}$He-Em | 3.7 / 14.8 | 2.44±0.11 | [20, 21] |
| $^{6}$Li-Em | 3.7 / 22.2 | 3.83±0.13 | [23, 24] |
| $^{7}$Li-Em | 2.2 / 15.4 | 2.2±0.1 | [24, 25] |
| $^{12}$C-Em | 3.7 / 44.4 | 4.7±0.2 | [22] |
| $^{16}$O-Em | 3.7 / 59.2 | 5.31±0.30 | [26, 27] |
| $^{22}$Ne-EM | 3.3 / 72.6 | 6.27±0.10 | [28, 29] |
| $^{24}$Mg-Em | 3.7 / 88.8 | 6.11±0.26 | [11] |
| $^{28}$Si-Em | 3.7 / 103.6 | 6.5±0.28 | [17] |
| $^{32}$S-Em | 3.7 / 118.4 | - | [20] |
| $^{56}$Fe-Em | 1.8 / 100.8 | 7.5±0.7 | Present work |
| $^{84}$Kr-Em | 1.0 / 84.0 | 7.31±0.16 | Present work |
| $^{132}$Xe-Em | 1.0 / 132.0 | 12.8±0.67 | Present work |

**Table 3:** The calculated values of wounded projectile nucleon ($W_P$), target nucleon ($W_T$), total number of interactions (ν) and produced average multiplicities of shower particles $<N_s>$ for different projectiles with different energies are listed.

| System | Energy (A GeV) | $<N_s>$ | $W_P$ | $W_T$ | ν | Reference |
|---|---|---|---|---|---|---|
| $^{24}$Mg-H | 3.7 | 2.02±0.15 | 1.75 | 1.00 | 1.75 | [11] |
| $^{28}$Si-H | 3.7 | 2.08±0.15 | 1.81 | 1.00 | 1.81 | [17] |
| $^{84}$Kr-H | 1.0 | 3.99±0.67 | 2.82 | 0.81 | 2.38 | Present work |
| P-CNO | 3.7 | 1.68±0.03 | 1.00 | 1.78 | 1.78 | [19] |
| $^{6}$Li-CNO | 3.7 | 3.58±0.16 | 2.78 | 3.53 | 4.38 | [23, 24] |
| $^{24}$Mg-CNO | 3.7 | 5.93±0.3 | 5.86 | 5.29 | 9.24 | [11] |
| $^{28}$Si-CNO | 3.7 | 6.11±0.28 | 6.34 | 5.53 | 9.99 | [17] |
| $^{84}$Kr-CNO | 1.0 | 8.10±0.90 | 10.23 | 6.18 | 18.18 | Present work |
| P-Ag/Br | 3.7 | 1.55±0.04 | 1.00 | 3.03 | 3.03 | [19] |
| $^{6}$Li-AgBr | 3.7 | 6.97±0.18 | 3.97 | 7.69 | 9.55 | [23, 24] |
| $^{24}$Mg-AgBr | 3.7 | 15.92±0.46 | 10.74 | 14.82 | 25.87 | [11] |
| $^{28}$Si-AgBr | 3.7 | 17.29±0.52 | 11.92 | 15.90 | 28.71 | [17] |
| $^{84}$Kr-AgBr | 1.0 | 15.09±0.60 | 21.07 | 21.26 | 62.49 | Present work |
| P-Em | 3.7 | 1.63±0.02 | 1.00 | 2.66 | 2.66 | [19] |
| $^{2}$H-Em | 3.7 | 2.50±0.1 | 1.57 | 3.29 | 3.32 | [20] |
| $^{3}$He-Em | 3.7 | 3.60±0.1 | 2.11 | 3.90 | 4.50 | [20, 21] |
| $^{4}$He-Em | 3.7 | 3.80±0.1 | 2.59 | 4.70 | 5.57 | [20, 21] |
| $^{6}$Li-Em | 3.7 | 5.56±0.17 | 3.40 | 5.88 | 7.30 | [23, 24] |
| $^{7}$Li-Em | 2.2 | 3.90±0.1 | 3.76 | 6.34 | 8.10 | [24, 25] |
| $^{12}$C-Em | 3.7 | 7.60±0.2 | 5.31 | 7.86 | 11.46 | [22] |
| $^{16}$O-Em | 3.7 | 8.80±0.5 | 6.31 | 8.74 | 13.69 | [26, 27] |
| $^{22}$Ne-Em | 3.3 | 10.50±0.1 | 7.66 | 9.67 | 16.60 | [28, 29] |
| $^{24}$Mg-Em | 3.7 | 11.12±0.38 | 8.06 | 10.00 | 17.46 | [11] |
| $^{28}$Si-Em | 3.7 | 11.64±0.35 | 8.76 | 10.51 | 18.98 | [17] |
| $^{32}$S-Em | 3.7 | 13.40±0.6 | 9.41 | 10.99 | 20.44 | [20] |
| $^{56}$Fe-Em | 1.8 | 10.53±0.68 | 13.27 | 12.77 | 33.49 | Present work |
| $^{84}$Kr-Em | 1.0 | 10.86±0.23 | 17.02 | 14.60 | 42.93 | Present work |
| $^{132}$Xe-Em | 1.0 | 17.40±0.70 | 22.29 | 16.85 | 56.24 | Present work |
| $^{197}$Au-Em | 1.0 | 16.43±3.43 | 28.15 | 19.01 | 71.01 | Present work |

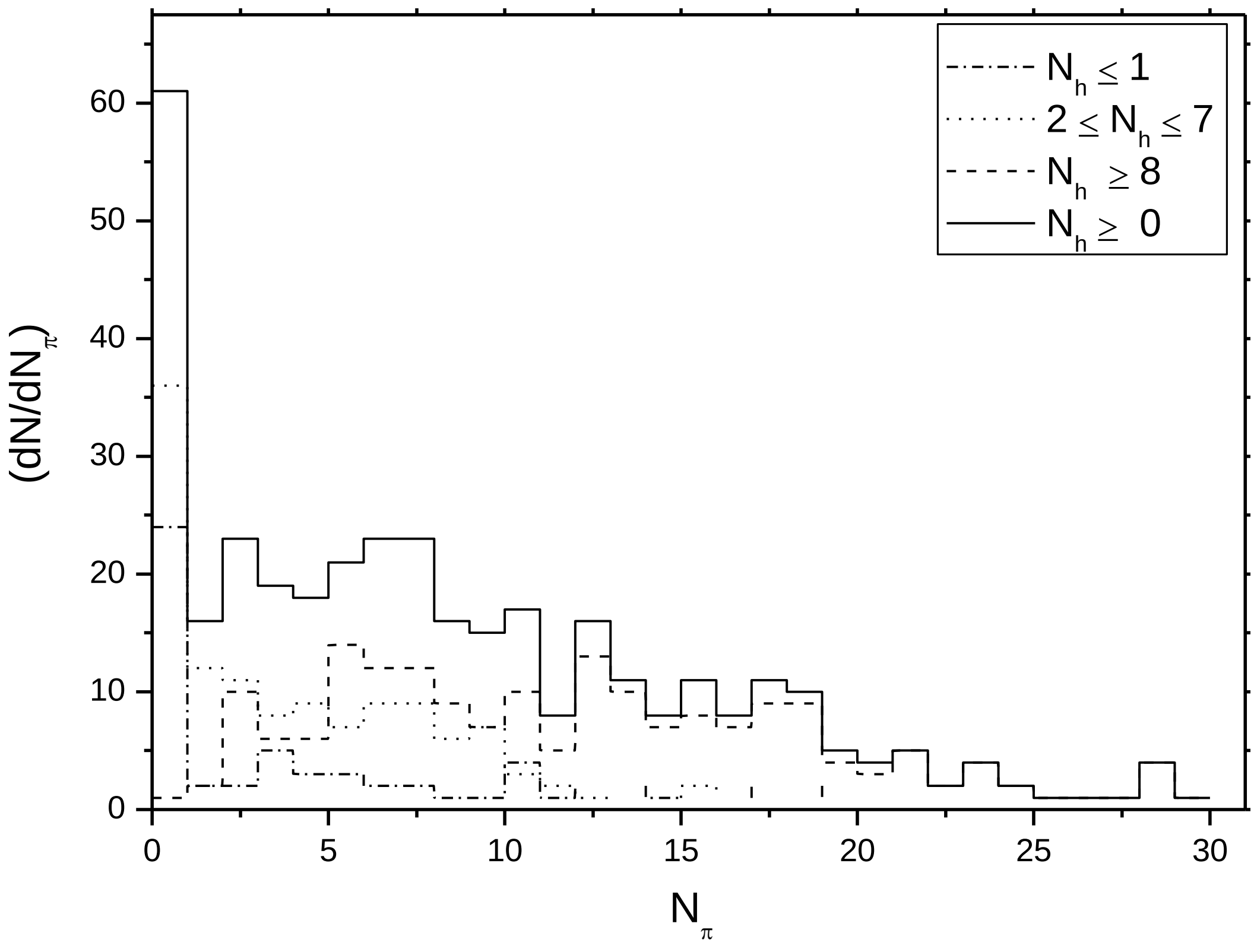

**Figure 1:** Multiplicity distribution of emitted charged pions ($N_\pi$) in the interaction of $^{84}Kr_{36}$ with nuclear emulsion target groups at around 1 A GeV.

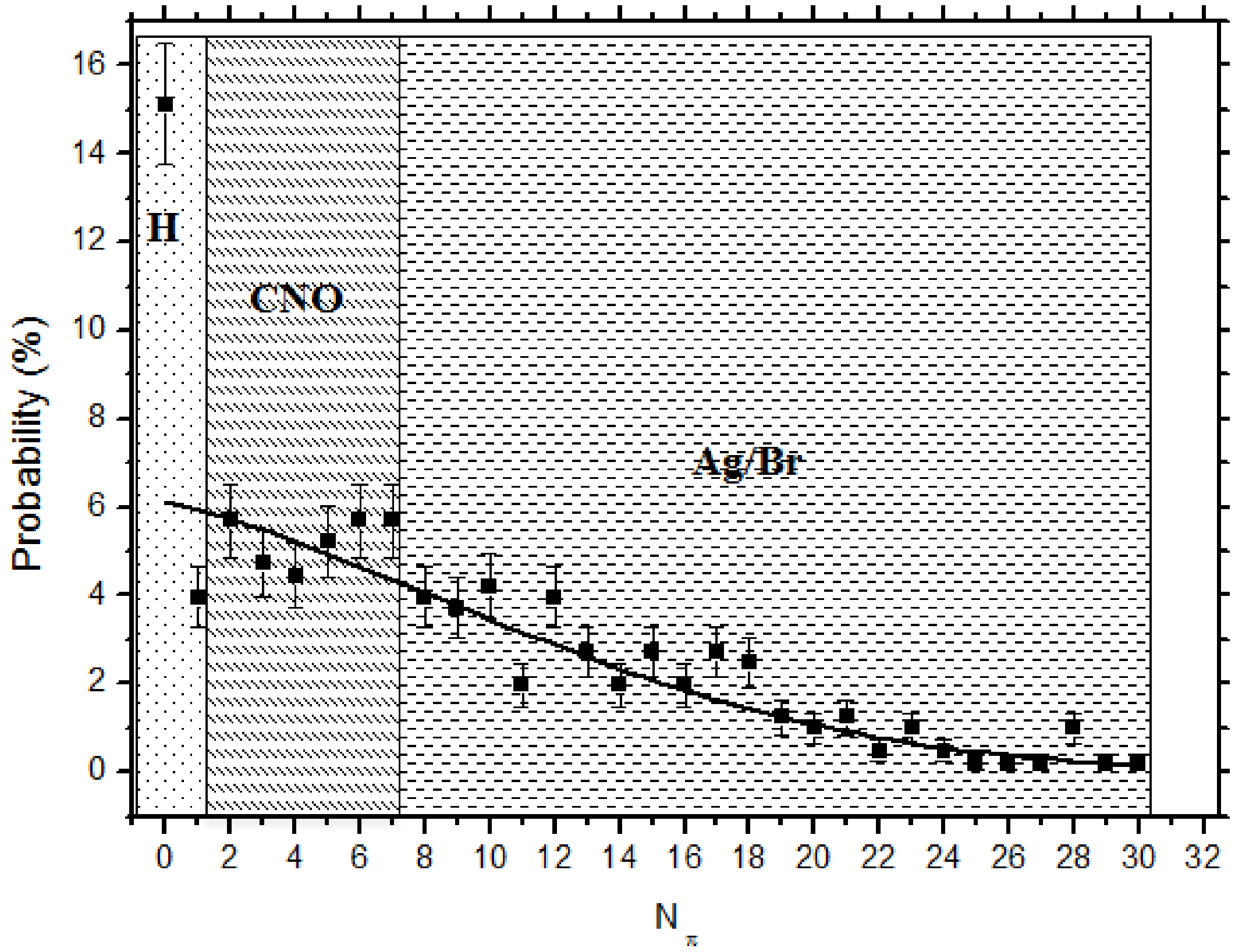

**Figure 2:** Probability distribution of emitted charged pions ($N_\pi$) in an interaction of $^{84}Kr_{36}$ – nuclei with emulsion nuclei at around 1 A GeV.

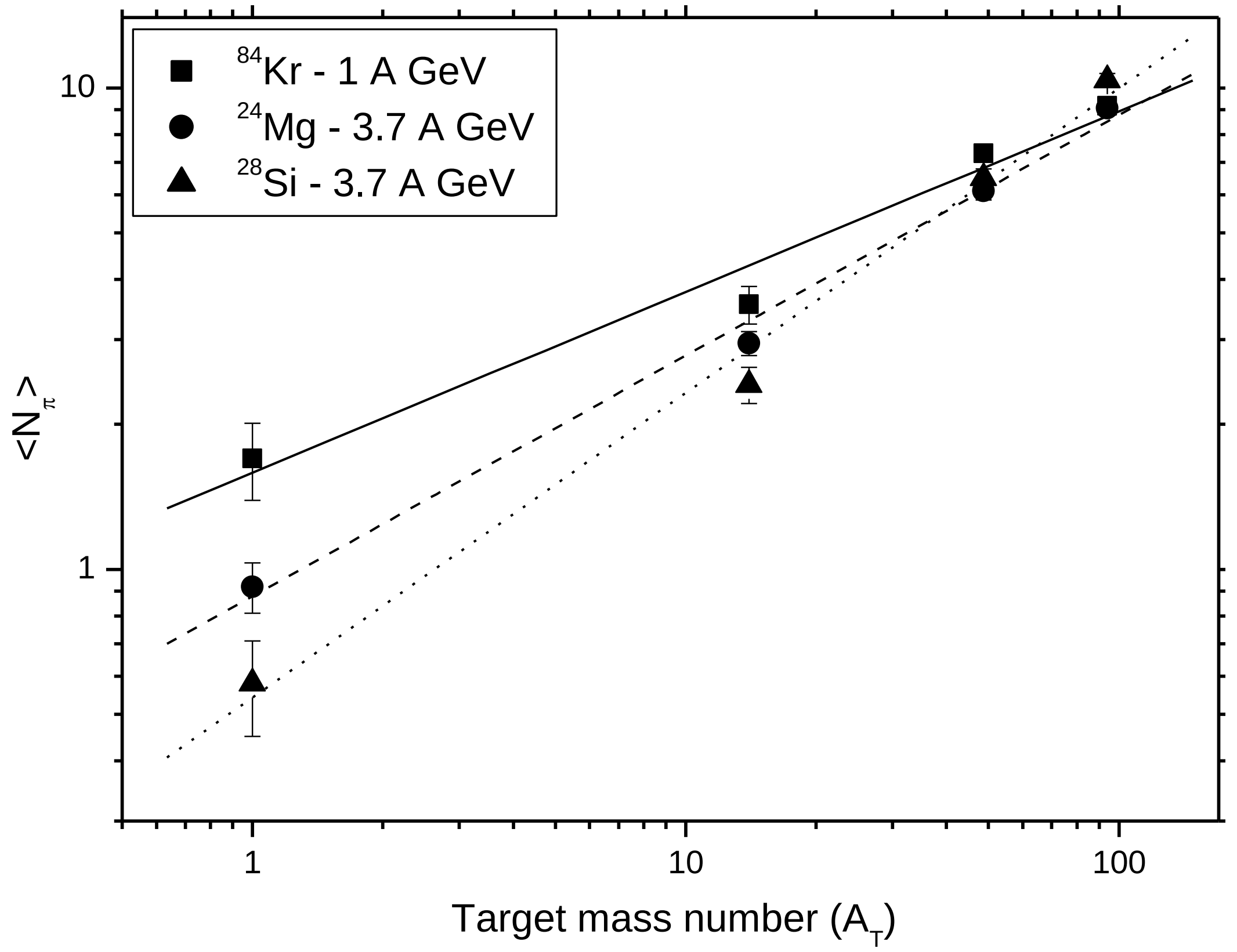

**Figure 3:** Dependence of average multiplicity of charged pion ($<N_\pi>$) as a function of target mass number ($A_T$) for different projectiles with different incident energy.

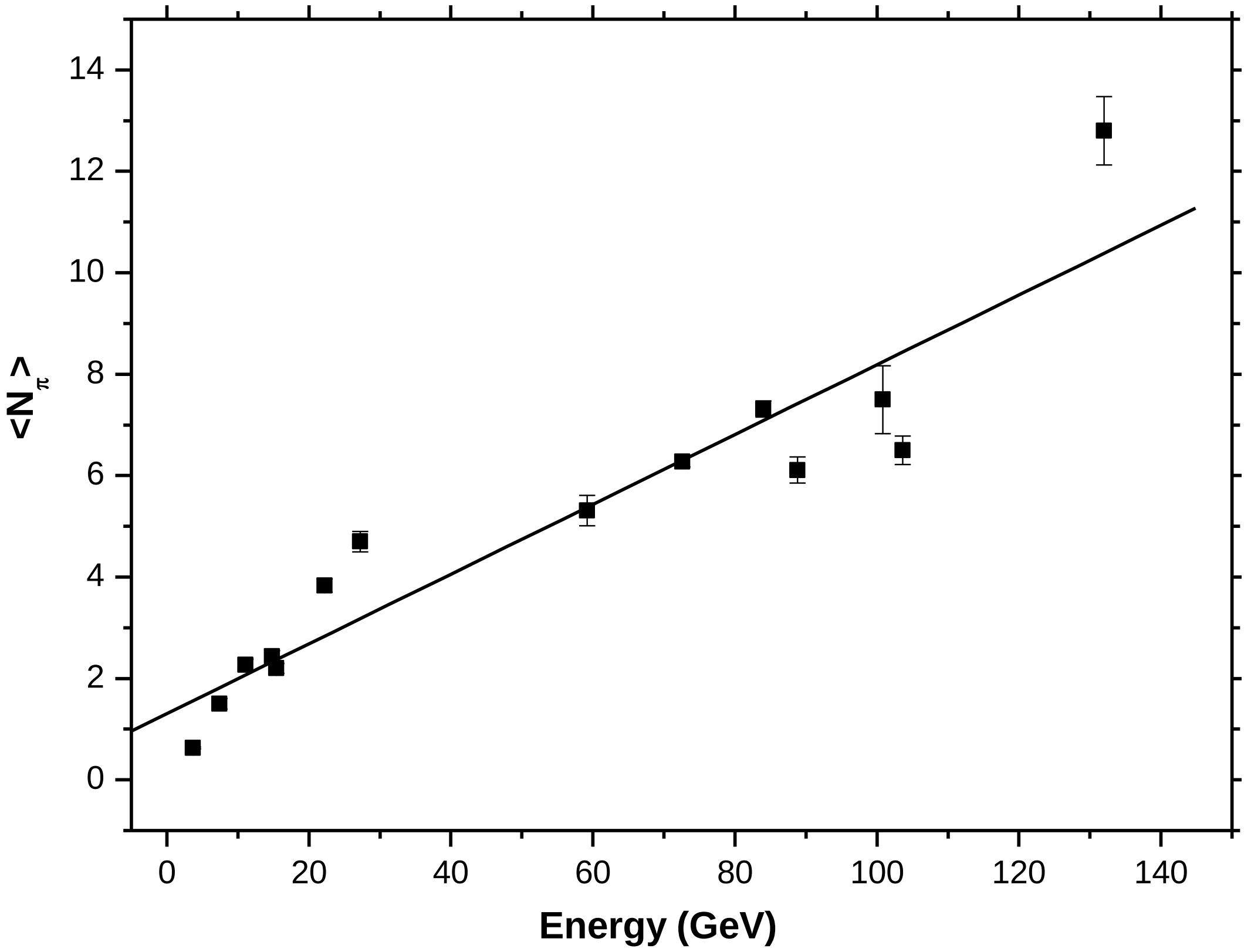

**Figure 4:** Correlation between average number of charged pions and total energy of the colliding projectiles.

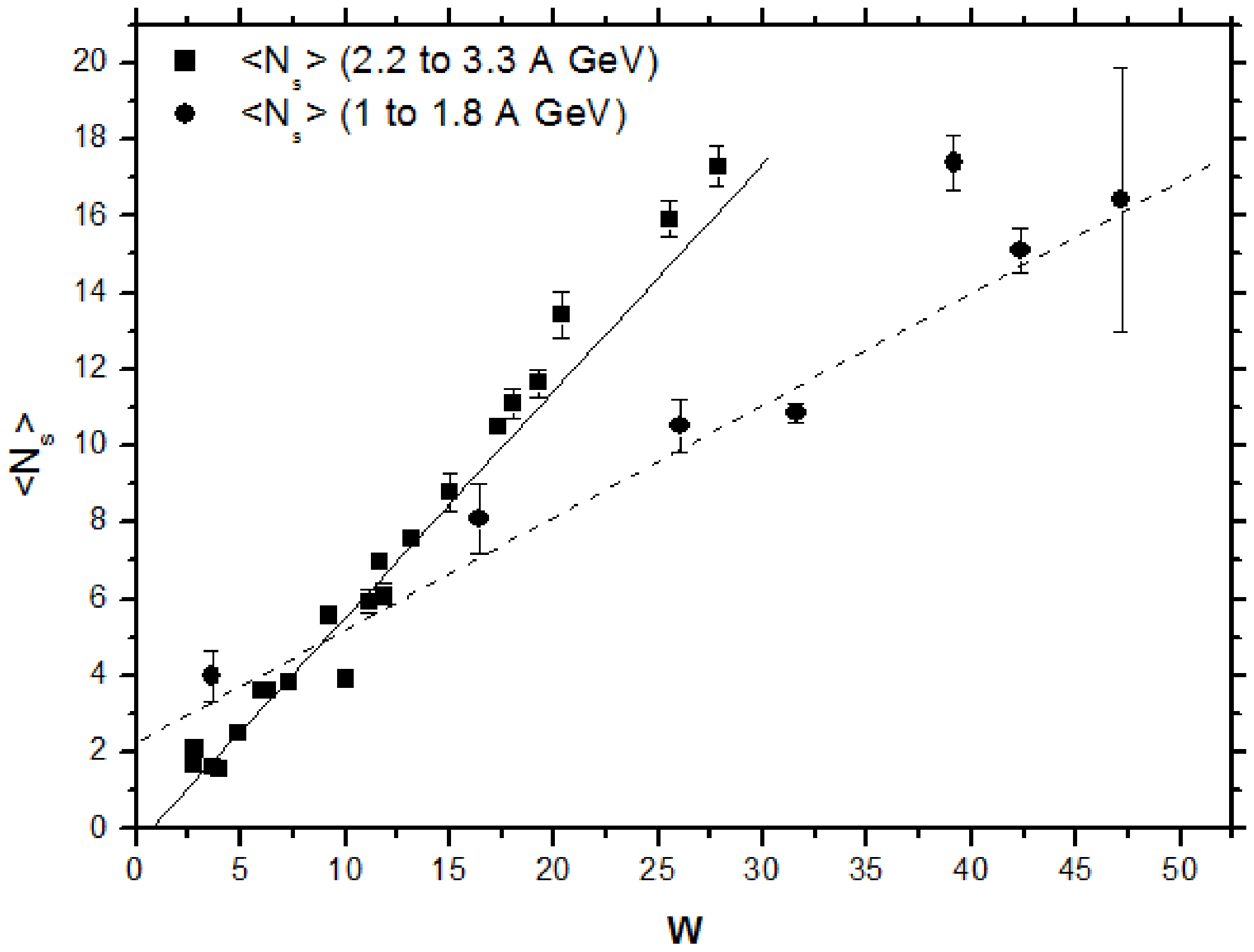

**Figure 5:** A relation between average numbers of produced shower particles <$N_s$> and estimated number of wounded nucleons (W) for different projectiles at different incident energies.

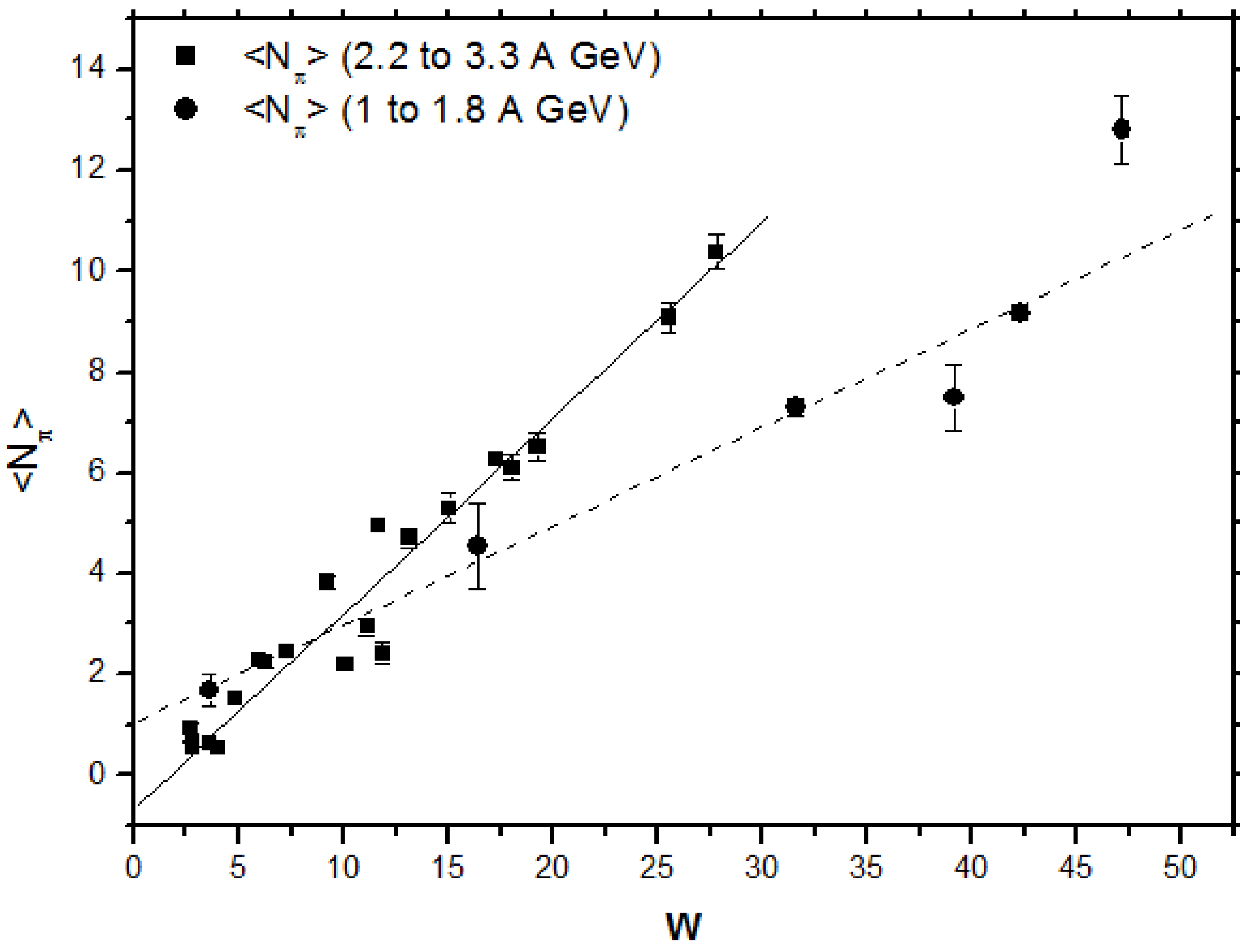

**Figure 6:** A relation between average numbers of produced pions <$N_\pi$> and the estimated number of wounded nucleons (W) for different projectiles at different kinetic energies.

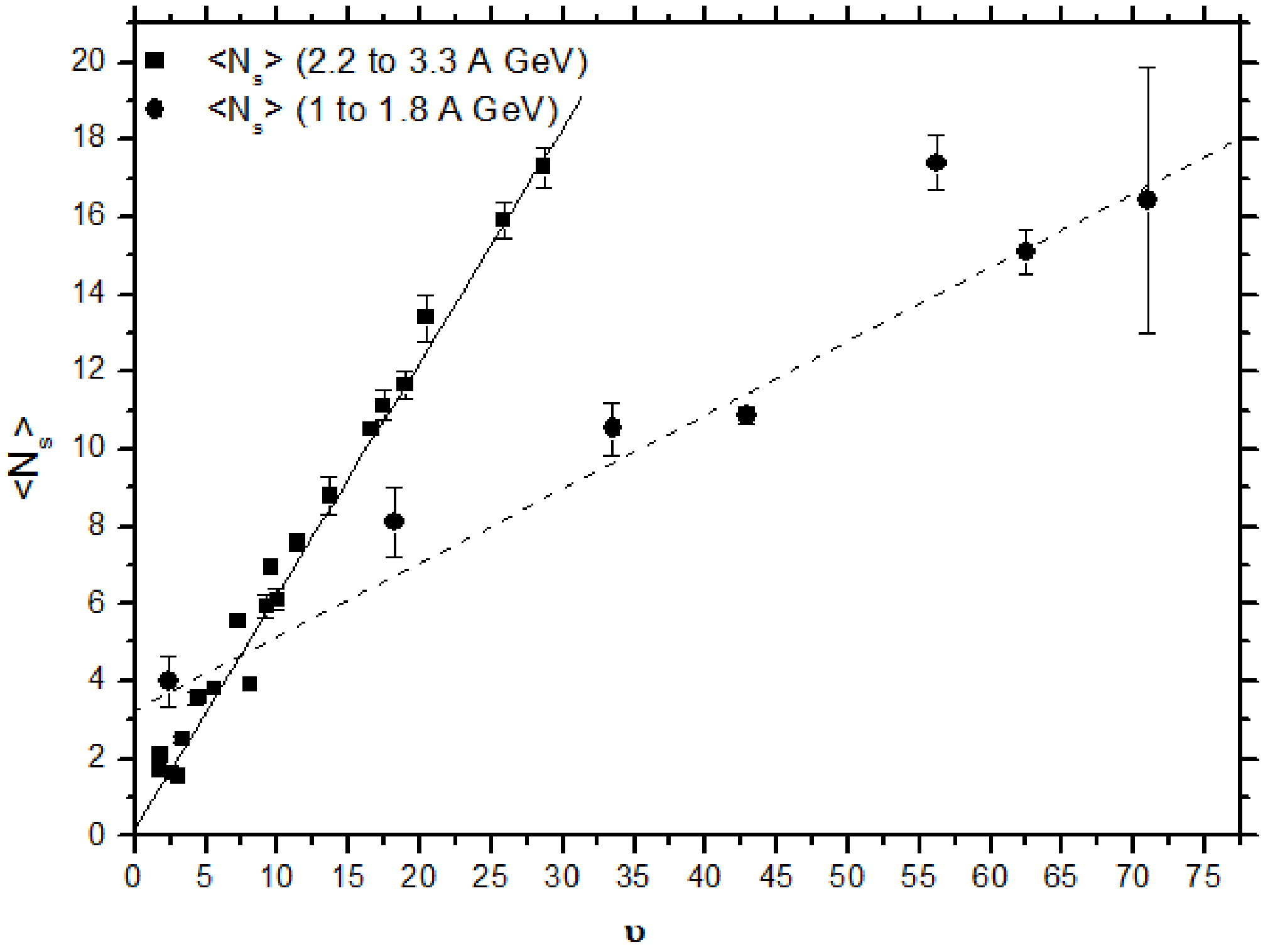

**Figure 7**: A dependence of the average numbers of produced shower particles <$N_s$> on the total number of estimated collisions (ν) for interactions of various projectiles with emulsion nuclei at different energies.

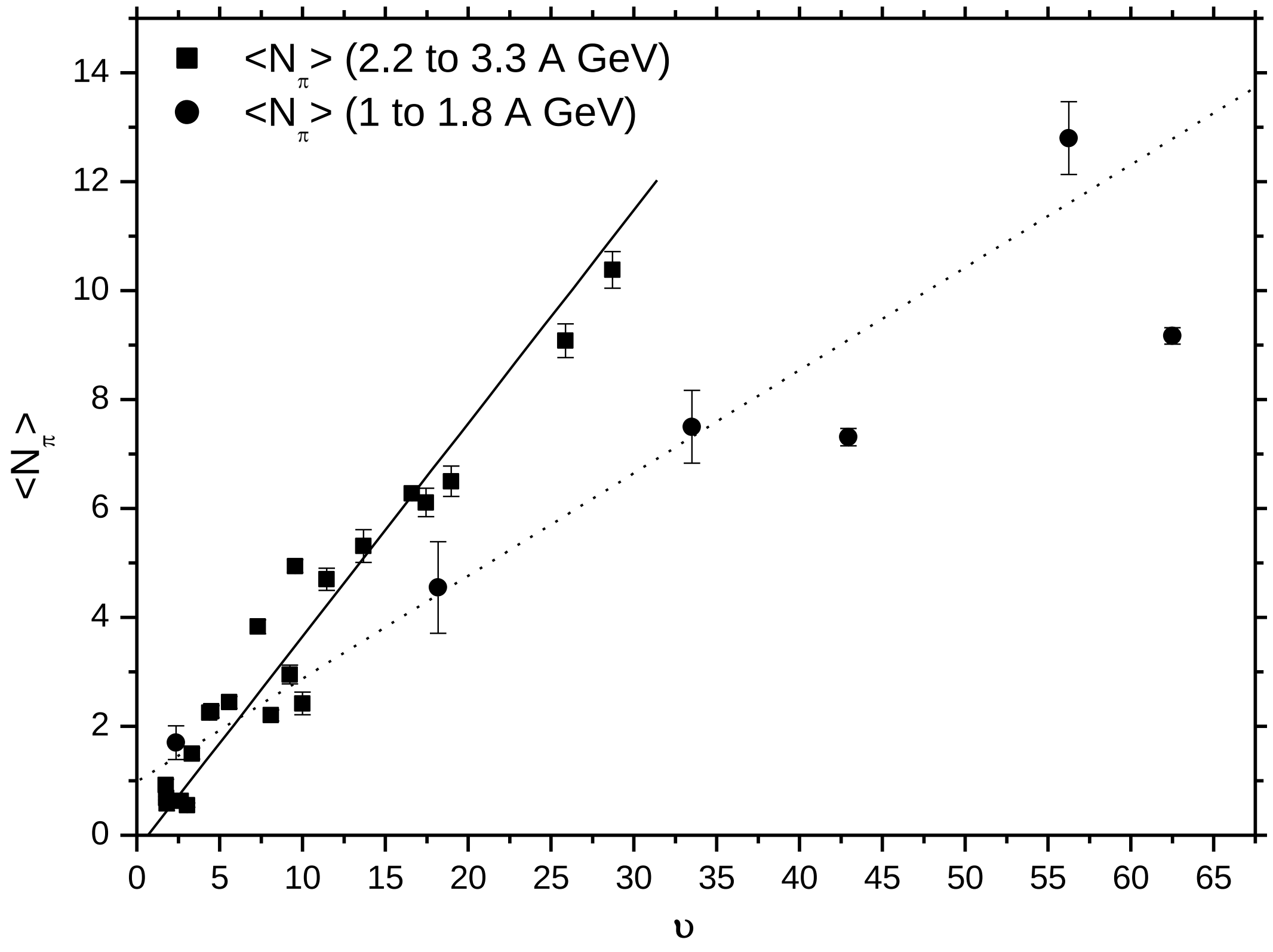

**Figure 8**: A dependence of the average numbers of produced pions particles <$N_\pi$> on the total number of estimated collisions (ν) for interactions of various projectiles with emulsion nuclei at different energies.

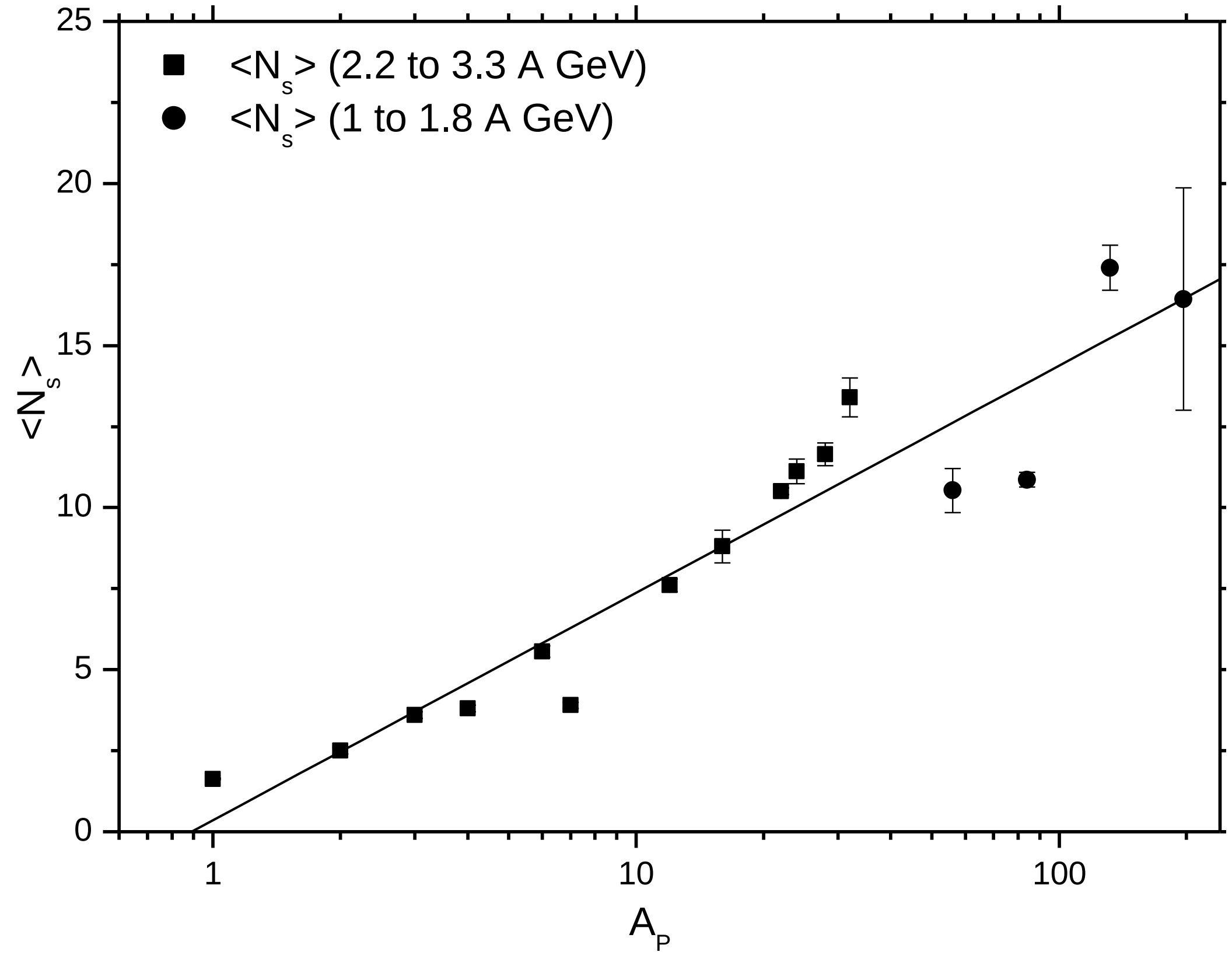

**Figure 9:** Dependence of average multiplicities of shower particles <$N_s$> on the projectile mass number ($A_P$) for inelastic collisions of various projectiles at different energy regions with emulsion nuclei.

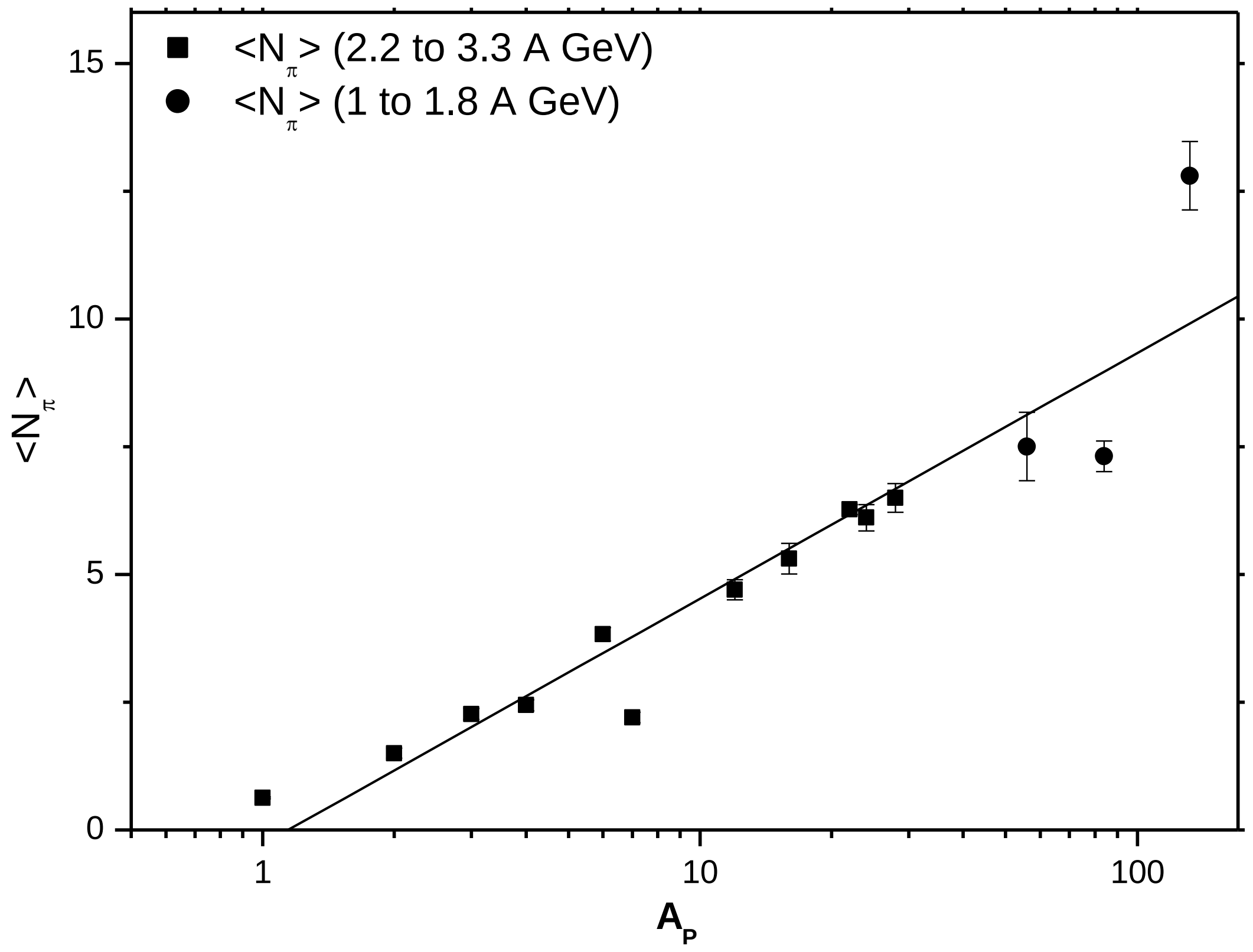

**Figure 10:** Dependence of average multiplicities of pion particles <$N_\pi$> on the projectile mass number ($A_P$) for inelastic collisions of various projectiles at different energy regions with emulsion nuclei.